\documentclass[aps,twocolumn,preprintnumbers,amsmath,amssymb]{revtex4}
\usepackage{epsfig}
\usepackage{amsmath}
\usepackage{graphicx}
\usepackage{dcolumn}
\usepackage{bm}
\usepackage{amssymb}
\usepackage{amsmath}
\usepackage{epsf}
\usepackage{subfigure}
\usepackage{epstopdf}
\usepackage{wrapfig}
\usepackage{pifont}
\usepackage{color}
\usepackage{wasysym}

\begin{document}

\newcommand{\pr}{\partial}
\newcommand{\rta}{\rightarrow}
\newcommand{\lta}{\leftarrow}
\newcommand{\ep}{\epsilon}
\newcommand{\ve}{\varepsilon}
\newcommand{\p}{\prime}
\newcommand{\om}{\omega}
\newcommand{\ra}{\rangle}
\newcommand{\la}{\langle}
\newcommand{\td}{\tilde}

\newcommand{\mo}{\mathcal{O}}
\newcommand{\ml}{\mathcal{L}}
\newcommand{\mathp}{\mathcal{P}}
\newcommand{\mq}{\mathcal{Q}}
\newcommand{\ms}{\mathcal{S}}

\newcommand{\nl}{$\newline$}
\newcommand{\nll}{$\newline\newline$}

\newcommand{\vspa}{\vspace{2mm}}
\newcommand{\vspb}{\vspace{3mm}}
\newcommand{\vspc}{\vspace{4mm}}

\title{A toy model for the pseudogap state of cuprates}

\author{Navinder Singh}
\affiliation{Physical Research Laboratory, Navrangpura, Ahmedabad-380009 India.}
\email{navinder.phy@gmail.com}

\begin{abstract}
A toy model is developed to capture "dance of pre-formed pairs" in the pseudogap state of high temperature cuprate superconductors. The model is a crude description of a hypothetical situation in which spin-driven spatial organization of the pre-formed pairs is analyzed. Within a mean-field theory (a slight generalization of Bragg-Williams theory) we examined the behaviour of heat capacity and an "order parameter" which captures the short range correlations between the pre-formed pairs. Below a transition temperature these short range correlations enhance and leads to a long range order (spin-density wave). This is examined through the breakdown of statistical independence.
\end{abstract}

\maketitle

\vspace{5mm}

Model building in physics is a first step towards understanding real complex physical systems\cite{peierls}. Here, we develop an Ising type model and present its mean-field solution under Bragg-Williams approximation. This toy model gets inspiration from the pre-formed pair picture of the pseudogap state of high temperature superconducting cuprates\cite{cup}. In under-doped cuprates, as one lowers temperature (below some characteristic temperature $T^\ast$), one gets a phase in which system is not superconducting, but shows a quasi-particle gap (in specific directions of the Brillouin zone). This non-superconducting state with quasi-particle gap is called the pseudogap state (recall that in simple metals (i.e., describable by BCS theory) quasi-particle gap and long-range superconducting order sets in at the same temperature called condensation temperature $T_c$). This gaped state in cuprates (which sets at $T^\ast>T_c$) without long range superconducting order is quite an elusive state\cite{elu}. There are many views on its origin\cite{ps,ps2}. In one prominent view this electronic state(phase) is pictured as a  system of pre-formed pairs  but without long-range superconducting order. As one lowers the temperature below $T_c$ these pre-formed pairs condense to give d-wave superconducting order(see for details\cite{ps}). 

\begin{figure}
\includegraphics[height = 5.5cm, width = 6cm]{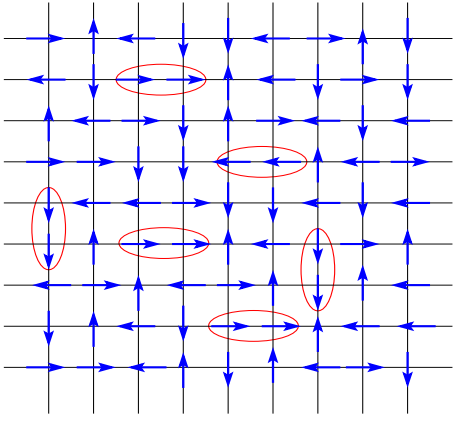}
\caption{A cartoon showing interacting pre-formed pairs. A pair is depicted by an arrow. Notation is as follows. At an arrowhead the quantum mechanical expectation value of a doped hole is pointing in the $+$ve z-direction (say, the z-direction is perpendicular to the plane of this paper) and at a tail it is in $-$ve z-direction.}
\label{main1}
\end{figure}

Here we develop a very crude model of the pseudogap state in which we assume that pre-formed pairs do exist and they interact with each other (figure~(\ref{main1}) shows one realization from an ensemble). We will see from a mean-field type solution that below  a temperature $T_0$ (a function of the parameters of the model) some sort of long range order is established. This is analyzed in detail in the following paragraphs. The phase transition in this model is the result of a competition between thermal agitation and pair alignment due to anti-ferromagnetic exchange interaction. Virtual hoping favors an arrangement in which all the pairs align in one direction. In this case adjacent spins in each pair (depicted by an arrow in figure~(\ref{main1})) are anti-parallel to each other and forms a low energy configuration due to exchange coupling mediated by virtual hoping. Thermal effects disrupt this and one has an ensemble of realizations (one is depicted in figure~(\ref{main1})).

{\it We state at the outset that this picturization of the pseudogap state is just a very crude description of the actual state of affairs. Actual physical picture is much more complex and elusive\cite{elu,ps,ps2}. {\bf Thus our interest here is purely of mathematical curiosity} (we enjoyed a mean-field type solution with slight generalization of the Bragg-Williams approximation).}

Now we develop a mathematical formulation of the picture. Let $N_R,~N_L,~N_U,$ and $N_D$ be the total number of pairs (arrows) pointing towards right, left, up, and down respectively. The total number of pairs is $N= N_R+N_L+N_U+N_D$. Next comes the energetics which are governed by pair-pair  interactions:
\begin{eqnarray}
H=&-&\ep_1 (\underbrace{N_{RR}^x}_{\rightarrow \rightarrow} + \underbrace{N_{LL}^x}_{\leftarrow \leftarrow} + \underbrace{N_{UU}^y}_{\frac{\uparrow}{\uparrow}} + \underbrace{N_{DD}^y}_{\frac{\downarrow}{\downarrow}} )\nonumber\\
&-&\ep_2 (\underbrace{N_{UD}^x}_{\uparrow \downarrow} + \underbrace{N_{RL}^y}_{\frac{\rightarrow}{\leftarrow}} )\nonumber\\
&+&\ep_3 (\underbrace{N_{UU}^x}_{\uparrow \uparrow} + \underbrace{N_{DD}^x}_{\downarrow \downarrow} + \underbrace{N_{RR}^y}_{\frac{\rightarrow}{\rightarrow}} + \underbrace{N_{LL}^y}_{\frac{\leftarrow}{\leftarrow}} )\nonumber\\
&+&\ep_4 (\underbrace{N_{RL}^x}_{\rightarrow \leftarrow} + \underbrace{N_{UD}^y}_{\frac{\uparrow}{\downarrow}})\nonumber\\
&+& 0 (\rightarrow \uparrow + \rightarrow \downarrow + \leftarrow \uparrow +...)
\label{eqn1}
\end{eqnarray}
The first term in the above equation (Eq.~(\ref{eqn1})) gives the contribution to total energy due to pair-pair (arrow-arrow) interaction energy in which pairs (arrows) are directed in the same direction (right-right and left-left  along horizontal axis, and up-up and down-down along the vertical axis). Some examples are shown in Fig.~(\ref{main1}) (circled). This energy is $-\ep_1$ (per pair, and $\ep_1$ is a positive quantity). Similarly, the second term gives the energy contribution due to parallel arrows (up-down along the horizontal axis , and left-right along the vertical axis). We do not distinguish between $N_{UD}^x$ and $N_{DU}^x$ etc. The 3rd and 4th terms contribute positive energy as $N_{UU}^x$, $N_{RL}^x$ etc etc are energetically costly (anti-ferromagnetic exchange will not favor these configurations). We put the interaction energy of the pairs like $N_{RU}^x$, $N_{RD}^x$, $N_{LD}^x$ etc etc to zero, the last term in Eq.~(\ref{eqn1}) (i.e., arrows which are at right angles to each other).

To calculate the partition sum and thermodynamical properties for this system we need to express cross terms like $N_{UD}^x$, $N_{RL}^y$ etc  in terms of non-cross terms like in the first term in equation ~(\ref{eqn1}). To find these relations we do a construction generally done in the Ising model\cite{ker}. Let us start with an arrow pointing towards right. Draw two horizontal lines from it, one towards right of it and the other towards left. Repeat this procedure for all the pairs pointing towards right. Total number of lines drawn will be $2 N_R$ (two line per right pointing arrow). There will be two lines between two adjacent arrows both pointing right (in the horizontal direction) and one line between two adjacent arrows one pointing right and the other pointing left, or up, or down. Thus $2 N_R = 2 N_{RR}^x+N_{RL}^x +N_{RU}^x+N_{RD}^x$.  Similar construction can be done in the vertical direction and for other pairs. Thus we will have a system of simultaneous linear equations:
\begin{eqnarray}
2 N_R &=& 2 N_{RR}^x+N_{RL}^x +N_{RU}^x+N_{RD}^x\nonumber\\
2 N_R &=& 2 N_{RR}^y+N_{RL}^y +N_{RU}^y+N_{RD}^y\nonumber\\
2 N_L &=& 2 N_{LL}^x+N_{RL}^x +N_{LU}^x+N_{LD}^x\nonumber\\
2 N_L &=& 2 N_{LL}^y+N_{RL}^y +N_{LU}^y+N_{LD}^y\nonumber\\
2 N_U &=& 2 N_{UU}^x+N_{RU}^x +N_{LU}^x+N_{UD}^x\nonumber\\
2 N_U &=& 2 N_{UU}^y+N_{RU}^y +N_{LU}^y+N_{UD}^y\nonumber\\
2 N_D &=& 2 N_{DD}^x+N_{RD}^x +N_{LD}^x+N_{UD}^x\nonumber\\
2 N_D &=& 2 N_{DD}^y+N_{RD}^y +N_{LD}^y+N_{UD}^y
\label{eqn2}
\end{eqnarray} 
We need to express cross terms ($2$nd and $4$th lines in equation (\ref{eqn1})) in termrs of non-cross terms like in the $1$st and $3$rd lines of equation (\ref{eqn1}). From the above array (equation (\ref{eqn2})) it is clear that we need to know $12$ variables ($N_{RL}^x, N_{RU}^x, N_{RD}^x, N_{RU}^y, N_{RD}^y...$ etc) and we only have eight equations. System is under-determined. For the solution, we put a reasonable approximation. Clearly the following configurations
\begin{equation}
\underbrace{\rightarrow \leftarrow}_{N_{RL}^x}~~{\rm and}~~\underbrace{\frac{\uparrow}{\downarrow}}_{N_{UD}^y}
\end{equation}
costs lots of energy (being disfavored by anti-ferromagnetic exchange). We project out these configurations from our ensemble of realizations. For this, we restrict that $k_B T << \ep_4$ (where $k_B$ is the Boltzmann's constant and $T$ is the temperature). Thus, in statistical realizations in an ensemble these configurations will rarely appear, and we can assume $N_{RL}^x$ and $N_{UD}^y$ $\simeq 0$. This projection reduces the unknown variables to $10$, while the system remains under-determined it is possible to write the second term in equation (\ref{eqn1}) in terms of non-cross variables. With simple algebra from equations (2):
\begin{eqnarray}
N_{UD}^x+N_{RL}^y&=&N_{RR}^x+N_{LL}^x+N_{UU}^y+N_{DD}^y\nonumber\\
&-&N_{UU}^x-N_{DD}^x-N_{RR}^y-N_{LL}^y.
\end{eqnarray}
With this, the Hamiltonian takes the form:
\begin{eqnarray}
&&H(\ep_1,\ep_2,\ep_3;N_{RR}^x,...)=\nonumber\\
&-&(\ep_1+\ep_2)(N_{RR}^x + N_{LL}^x + N_{UU}^y + N_{DD}^y)\nonumber\\
&+&(\ep_2 +\ep_3) (N_{RR}^y + N_{LL}^y +N_{UU}^x + N_{DD}^x.
\label{eqn4}
\end{eqnarray}
The partition function can be calculated by summing over all possible values that these pairs can take (with constraints). Formally
\begin{widetext}
\begin{equation}
Z(\ep_1,\ep_2,\ep_3;T)=\underbrace{\sum_{N_{RR}^x}\sum_{N_{LL}^x}\sum_{N_{UU}^x}\sum_{N_{DD}^x}\sum_{N_{RR}^y}\sum_{N_{LL}^y}\sum_{N_{UU}^y}\sum_{N_{DD}^y}}_{{\rm with~~ constraints}} W(N_{RR}^x,N_{LL}^x,N_{UU}^x...) e^{-\beta H(\ep_1,\ep_2,\ep_3;N_{RR}^x,N_{LL}^x,N_{UU}^x,...)}
\end{equation}
\end{widetext}
$\beta =\frac{1}{k_B T}$. To know the constraints and the statistical weight ($W$) for given numbers of $N_{RR}^x...etc$ is a formidable task as these non-cross numbers are coupled through cross numbers dictated by the geometry of the model. Here we do a simple mean-field approximation; a slight generalization of the Bragg-Williams approximation. We assume that (as done in Bragg-Williams approximation\cite{ker,bragg}) the number of pairs (near-neighbor) of right pointing arrows is proportional to the square of the total number of right pointing arrows:
\begin{equation}
\frac{N_{RR}}{2 N} = \frac{N_{RR}^x}{2 N} + \frac{N_{RR}^y}{2N} = \left(\frac{N_R}{N}\right)^2
\end{equation}
The meaning of this is that "we do not have short range order apart from that which comes from long-range order" (see for details\cite{ker,bragg}). We further assume that:
\begin{equation}
\frac{N_{RR}^x}{2N} = \alpha \left(\frac{N_R}{N}\right)^2
\end{equation}
Here $\alpha$ varies from $0$ to $1$ (ensemble contains all realizations of the mixture of $N_{RR}^x$ and $N_{RR}^y$). Thus
\begin{equation}
\frac{N_{RR}^x}{N_{RR}^y} = \frac{\alpha}{1-\alpha}
\end{equation}
On similar lines we have $\frac{N_{LL}^x}{2N} = \beta \left(\frac{N_L}{N}\right)^2$, $\frac{N_{UU}^x}{2N} = \gamma \left(\frac{N_U}{N}\right)^2$, and $\frac{N_{DD}^x}{2N} = \delta \left(\frac{N_D}{N}\right)^2$. By writing $N_{RR}^x$ etc in terms of $N_R$ etc the Hamiltonian takes the following form:
\begin{eqnarray}
H(\ep_a,\ep_b;\alpha,..;N_R,..)&=& 2 N (N_R/N)^2 (\ep_b-(\ep_a+\ep_b)\alpha)\nonumber\\
&+& 2 N (N_L/N)^2 (\ep_b-(\ep_a+\ep_b)\beta)\nonumber\\
&+& 2 N (N_U/N)^2 (\ep_b-(\ep_a+\ep_b)\gamma)\nonumber\\
&+& 2 N (\frac{N_D}{N})^2 (\ep_b-(\ep_a+\ep_b)\delta)
\end{eqnarray}
Where $N_D = 1-\frac{N_R+N_L+N_U}{N}$ (as $N = N_R+N_L+N_U+N_D$). And we define $\ep_a =\ep_1+\ep_2,~~\ep_b=\ep_2+\ep_3$. After this mean-field approximation the partition function can simply be written as:

\begin{widetext}
\begin{equation}
Z(\ep_a,\ep_b;T)=\sum_{N_U=0}^{N}\sum_{N_L=0}^{N-N_U}\sum_{N_R=0}^{N-N_U-N_L}\int_0^1 d\alpha \int_0^1 d\beta\int_0^1 d\gamma\int_0^1 d\delta  \frac{N!}{N_U!N_L!N_R!(N-N_U-N_L-N_R)!} e^{-\beta H(\ep_a,\ep_b;\alpha,..;N_R,..)}
\label{final1}
\end{equation}

Integrals can be performed easily and the partition function takes the form:

\begin{equation}
Z(\ep_a,\ep_b;T)=\frac{1}{(\ep_a+\ep_b)^4}\sum_{N_U=0}^{N}\sum_{N_L=0}^{N-N_U}\sum_{N_R=0}^{N-N_U-N_L} \frac{N!}{N_U!N_L!N_R!(N-N_U-N_L-N_R)!}\prod_{i} \frac{e^{2 N\beta \ep_a (N_i/N)^2} -e^{2 N\beta \ep_b (N_i/N)^2}}{2 N \beta (N_i/N)^2}
\label{finall}
\end{equation}
\end{widetext}
Here $i$ takes the values $R, L, U$, and $D$, and $N_D/N$ can always be written as $1-(N_R+N_L+N_U)/N$. Free energy can be calculated from the fundamental relation of statistical thermodynamics: $F(\ep_a,\ep_b,T) = -\frac{1}{\beta} Log(Z(\ep_a,\ep_b;T))$, and other thermodynamical functions can be easily calculated: internal energy $U=-k_B T^2 \frac{\pr}{\pr T}\left(\frac{F}{k_B T}\right)$; heat capacity: $C=\frac{\pr U}{\pr T}$.

\begin{figure}[h!]
\includegraphics[height = 5cm, width = 9cm]{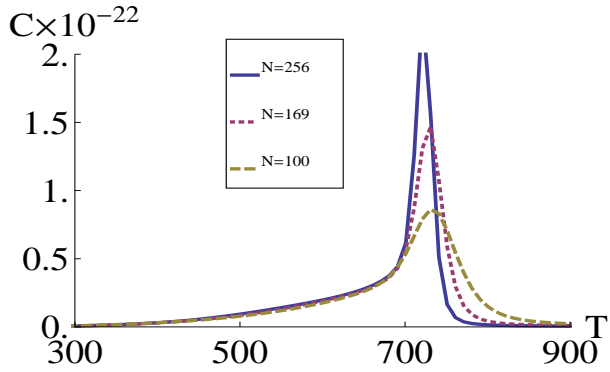}
\caption{Heat Capacity $C$ per pair (per arrow) in Joules per Kelvin as a function of temperature (in Kelvin) for various lattice sizes.  Below critical temperature $T_0= 731~K$ long range order sets in (as analysed below) and heat capacity becomes larger as compared to disordered phase. In the calculation we used $\ep_1=30 ~meV,~\ep_2=20~meV$, and $\ep_3=20~meV$.}
\label{heatcap1}
\end{figure}

In Fig.~(\ref{heatcap1}) we plot a numerical calculation of the heat capacity per pair (i.e., per arrow) as a function of temperature from equation (\ref{finall}). Three results are plotted for lattice sizes $N = 10\times10, ~13\times13$, and $16\times 16$. As the lattice size increases the peak in the heat capacity becomes sharper (in accord with the extended singularity conjuncture\cite{leo}).
The peak occurs at around $T_0\simeq 731~K$ (for $\ep_1=30~meV, \ep_2=\ep_3=20~meV$). Below this temperature some kind of long-range order sets in (as analyzed below). Heat capacity in the ordered phase $T<T_0$ is larger as compared to disordered phase $T>T_0$, as expected. We know from the Ginsburg's argument\footnote{Leo P. Kadanoff, {\it Statistical Physics}, World Scientific, Singapor (2000).} that at the critical point mean-field theory fails in $1$-D, 2-D, and in 3-D ({\it fluctuations dominate over the average behaviour}). Thus in the present case also, true behaviour at the critical point is not revealed  within the Bragg-Williams approximation. One has to go beyond it (for example by using the renormalization group analysis\footnote{{\it Phase transitions and critical phenomena}, Ed. C. Domb, and M. S. Green, Vol. 6, Academic Press, London (1976).}).

We will see from the fluctuation behaviour of an "order parameter" that the ordered state is a highly correlated state. This means that the statistical independence ($ RF=\frac{\sqrt{<f^2>-<f>^2}}{<f>}\propto \frac{1}{\sqrt{N}}$~, where $f$ is some "order parameter") is violated. Above the ordering temperature ($T>T_0$), statistical correlations break, and various parts of the system behave independently which leads to the resort of statistical independence. Below the ordering temperature $T<T_0$ we get highly fluctuating phase in which the Relative Fluctuation (RF) is greater than one! (rms fluctuations greater than the mean value). This seems to be independent from the lattice size as analysed below. This is a peculiarity of this phase in this toy model and may have analogies to the anomalous behaviour of the pseudogap state of cuprates (i.e., marked deviations from Fermi liquid behaviour which points towards the breakdown of adiabatic continuity\cite{and}). We again stress that this is just a speculation (as no order parameter has been discovered in the pseudogap state of cuprates).

\begin{figure}[h!]
\includegraphics[height = 5cm, width = 9cm]{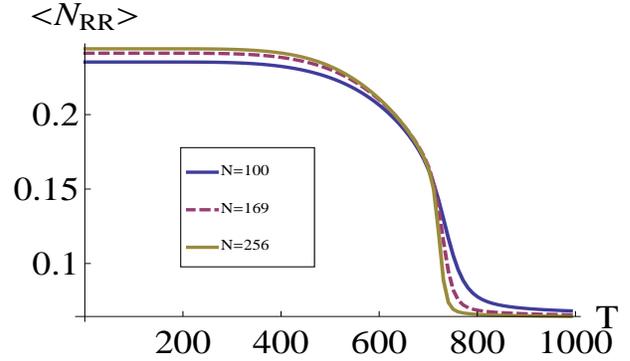}
\caption{Fraction of the average number of right pointing arrow pairs $\frac{<N_{RR}>}{2 N}$ as a function of temperature. The model parameters are: $\ep_1=30~meV$, $\ep_2=\ep_3=20~meV$.}
\label{fr}
\end{figure}

Next, we analyze the behaviour of an "order parameter" as the temperature is lowered through the transition temperature. We take $<N_{RR}>$ as an "order parameter". $<N_{RR}>$ is the ensemble average of the total number of two nearby right pointing arrows (arrow pairs) and $N_{RR} = N_{RR}^x (\rightarrow \rightarrow)+ N_{RR}^y (\frac{\rightarrow}{\rightarrow})$. We see from Fig.~(\ref{fr}) that in the ordered phase $T<T_0$ the average of the fraction $\frac{N_{RR}}{2 N}$ (ratio between $<N_{RR}>$ and the total number of all possible pairs $2N$) is about $1/4$.  
As the temperature is increased above $T_0$ this ratio falls, arrow pairs break (it is $0.06$ at $T=1000~K$; thus only $6$ percent of the total number of pairs are right pointing in this regime). In Fig.~(\ref{fr}) numerical results are plotted for three lattice sizes $10\times10,~13\times13,$ and $16\times 16$. The transition is sharper with larger lattice sizes (in accord with extended singularity theorem\cite{leo}). Thus in the ordered phase typical configurations has more $<N_{RR}>$ as compared to that in the disordered phase ($T>T_0$). From this numerical calculation we also notice that : $<N_{RR} -N_{LL}>=<N_{UU}-N_{DD}>=0$ and $<N_R>=<N_L>=<N_U>=<N_D>=N/4$ at all temperatures. This means that we have equal number of right and left pointing arrows and also equal number of arrow pairs (similarly for the up and down directions). Thus our "order parameter" $<N_{RR}>$ ($<N_{LL}>$ or $<N_{UU}>$ or $<N_{DD}>$ can also be used) do not induce directional inhomogeneity (more pairs in one direction and lesser in the opposite direction) but it captures the development of short-range pair correlations which leads to long range order in the ordered pahse $(T<T_0)$\footnote{that is why the word "order parameter" is written in quotes.} (as analyzed below). This is the object of our study.

As $<N_{RR}> = <N_{RR}^x>+<N_{RR}^y>$ we further analyze how $N_{RR}^x (\rightarrow \rightarrow)$ and $N_{RR}^y (\frac{\rightarrow}{\rightarrow})$ behave with temperature and system parameters. 
\begin{figure}[h!]
\centering
\begin{tabular}{cc}
\includegraphics[height = 3cm, width =4.3cm]{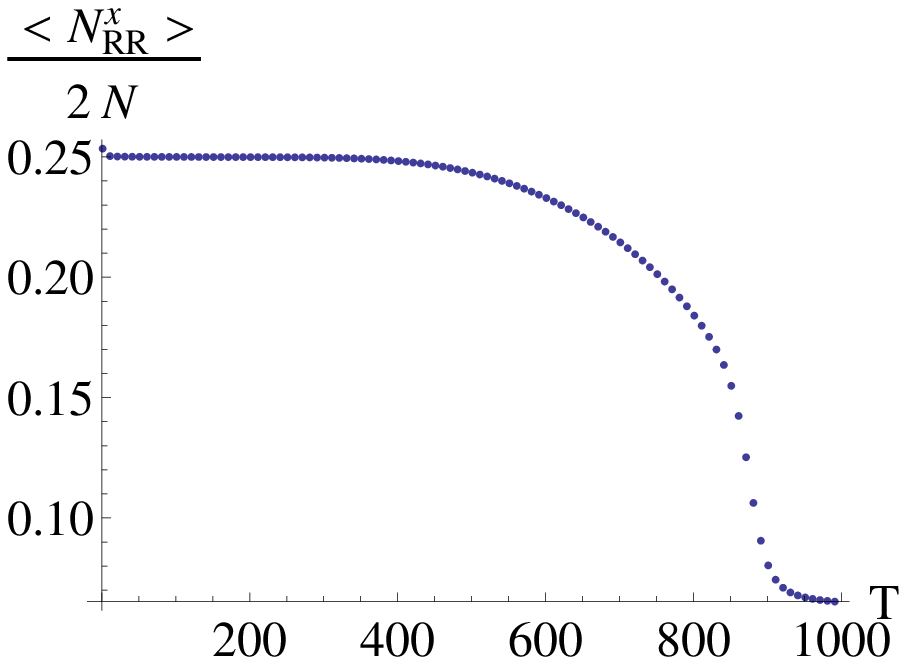}&
\includegraphics[height = 3cm, width =4.3cm]{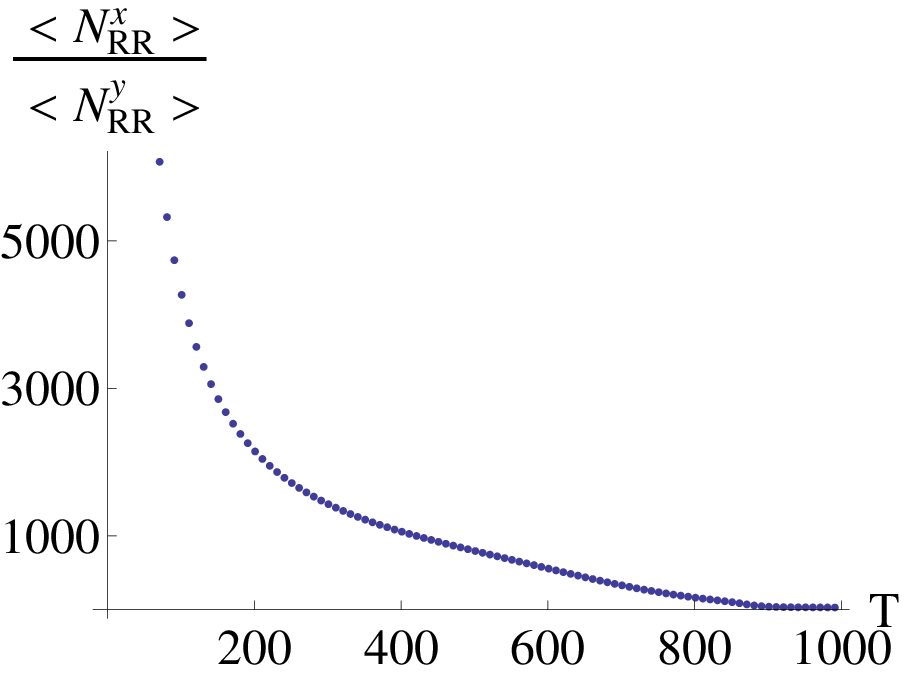}\\
(a)&
(b)\\
\includegraphics[height = 3cm, width =4.3cm]{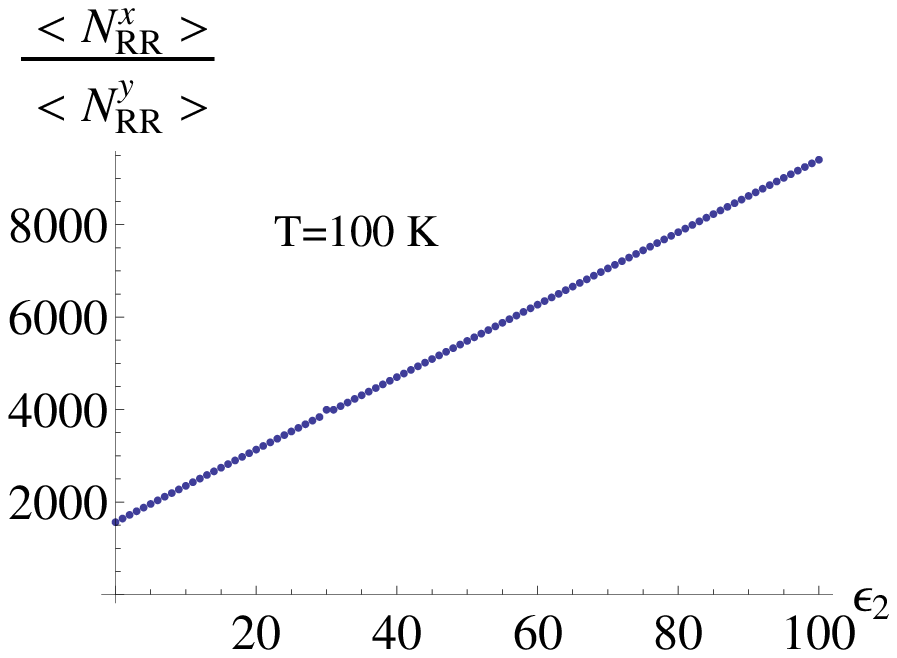}&
\includegraphics[height = 3cm, width =4.3cm]{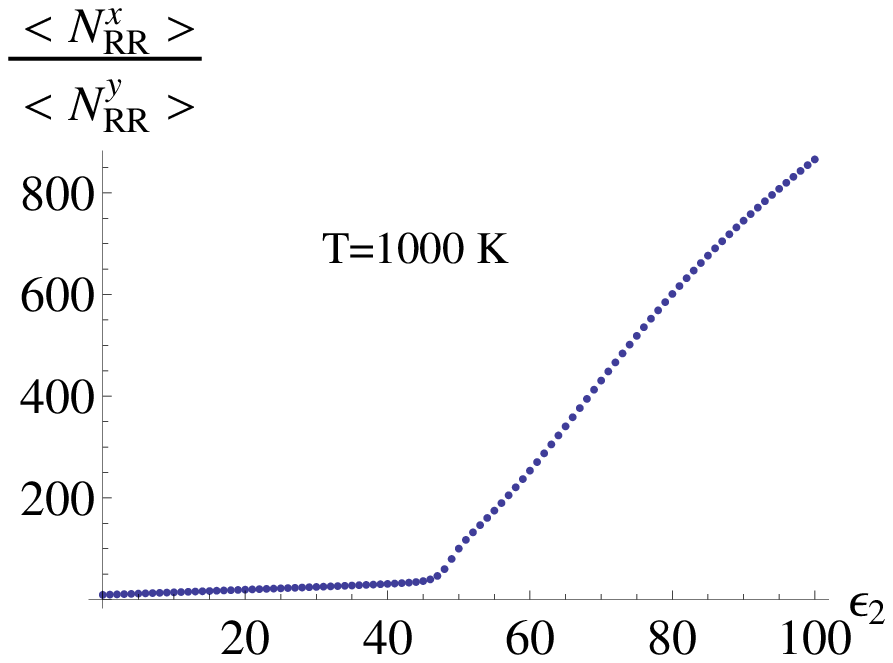}\\
(c)&
(d)\\
\end{tabular}
\caption{The behaviour of $<N_{RR}^x>$ and the ratio $\frac{<N_{RR}^x>}{<N_{RR}^y>}$ with temperature and with system parameter $(\ep_2)$.}
\label{xy}
\end{figure}
Figure (\ref{xy}(a)) shows how $\frac{<N_{RR}^x>}{2 N}$ changes with temperature (this behaviour is similar to that depicted in figure (\ref{fr})). From Fig.~(\ref{xy}(b)) we see that majority of the pairs are $N_{RR}^x$ (as $\frac{<N_{RR}^x>}{<N_{RR}^y>} \sim 10^3$). This is in accord with the input Hamiltonian. If we examine equation (\ref{eqn4}) we notice that the energy of $N_{RR}^x$ is negative and $N_{RR}^y$ pairs is positive (thus $\rightarrow \rightarrow$ are energetically favorable).  Figures (\ref{xy}(c)) and (\ref{xy}(d)) depictes how this ratio behave with $\ep_2$ (i.e., the negative energy of anti-parallel pairs). As expected, this ratio increases with $\ep_2$ (the parallel arrangement along the y-direction ($\frac{\rightarrow}{\rightarrow}$)  is energetically disfavored). The trend at low temperature ($T=100~K<T_0$) is linear but at high temperature ($T=1000~K >T_0$) there is a break in the slope (Fig.~\ref{xy}(d)). This is quite elusive. Why should the competition between $<N_{RR}^x>$ and $<N_{RR}^y>$ changes its nature at $\ep_2\simeq 45~meV$ (with $\ep_1=\ep_2=20~meV$)?

\begin{figure}[h!]
\includegraphics[height = 5cm, width = 9cm]{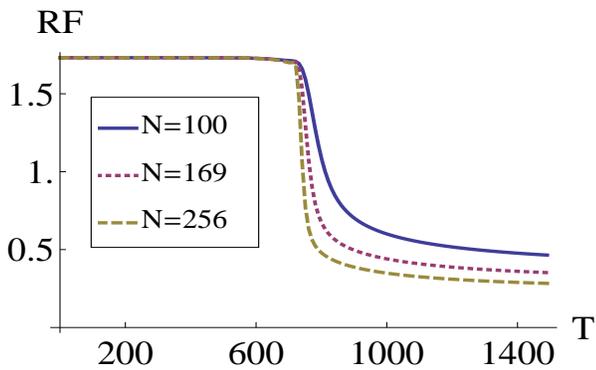}
\caption{Relative Fluctuation (RF) as a function of temperature for three lattice sizes as indicated in the legend.}
\label{rfl}
\end{figure}

Next we analyze the development of long range order. The fundamental thesis of Bragg-Williams approximation is: "we do not have short range order apart from that which comes from long-range order". To investigate the development of long-range order from this short ranged orientational order, as the temperature is lowered through $T_0$, we use statistical independence as a tool.  We study the quantity: $RF\equiv \frac{\sqrt{<N_{RR}^2>-<N_{RR}>^2}}{<N_{RR}>}$. This is the Relative Fluctuation (RF) and is plotted in Fig.~(\ref{rfl}) as a function of temperature for three lattice sizes as shown.
We see that $T<T_0$ phase is highly fluctuating (and roughly is lattice size independent). RF is $>1$ in this phase. At $T>T_0$ RF falls with increasing temperature (Fig.~(\ref{rfl})).

\begin{figure}[h!]
\centering
\begin{tabular}{cc}
\includegraphics[height = 3cm, width =4.3cm]{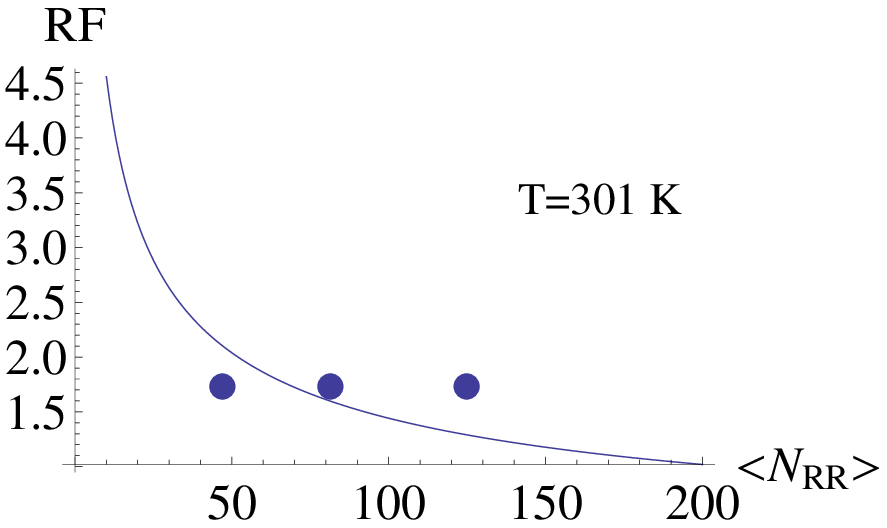}&
\includegraphics[height = 3cm, width =4.3cm]{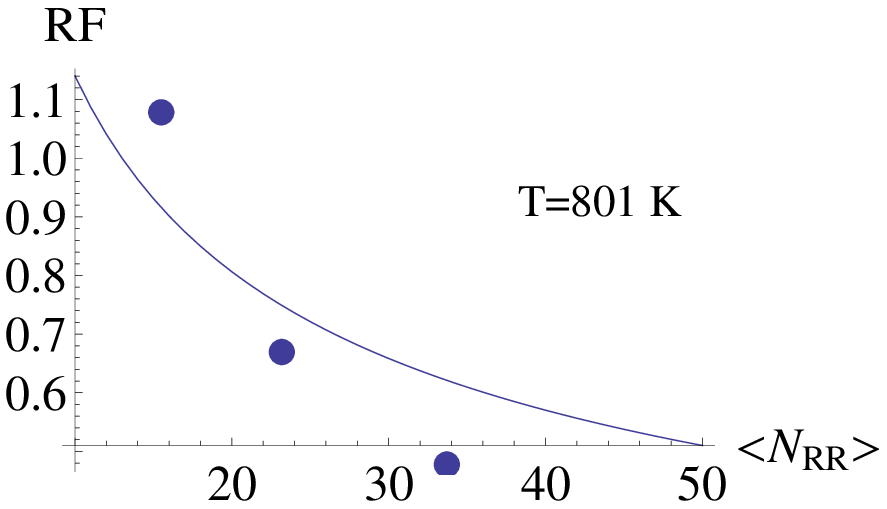}\\
(a)&
(b)\\
\includegraphics[height = 3cm, width =4.3cm]{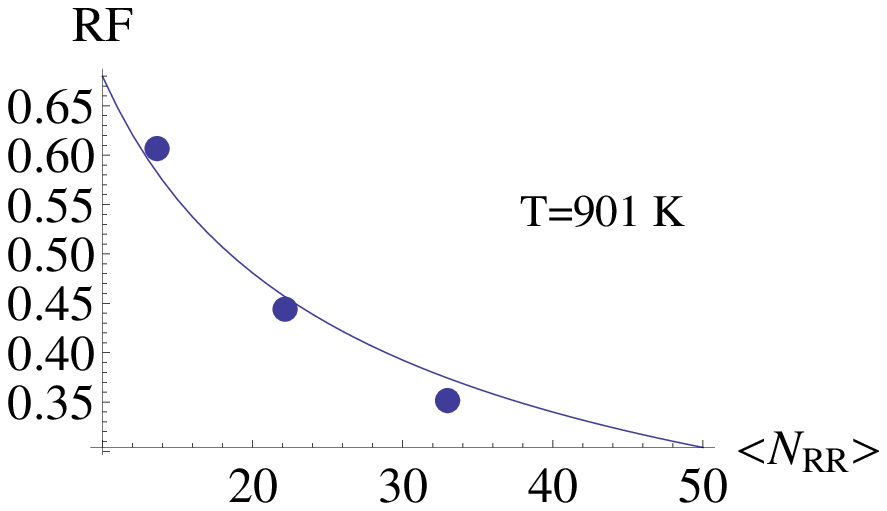}&
\includegraphics[height = 3cm, width =4.3cm]{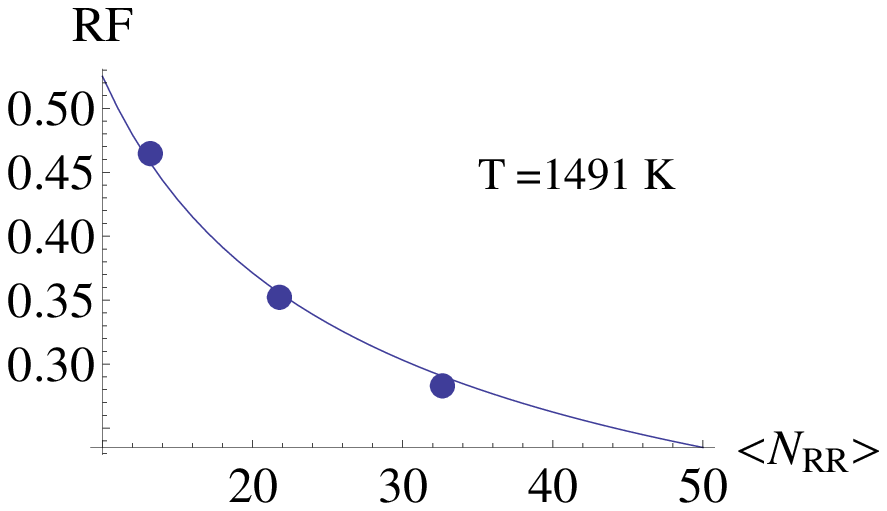}\\
(c)&
(d)\\
\end{tabular}
\caption{Filled circles are $RF(<N_{RR}>)=\frac{\sqrt{<N_{RR}^2>-<N_{RR}>^2}}{<N_{RR}>}$ for three lattice sizes (as mentioned before) and the line is the least square fit using $RF(<N_{RR}>)=\frac{a}{\sqrt{<N_{RR}>}}$ ($a$ is determined by least-square fitting).}
\label{st}
\end{figure}

For a system with short range interactions (like usual statistical mechanical systems, for example, gas in a box), relative fluctuation in some observable, say $f$, are proportional to $1/\sqrt{N}$:
\[\frac{\sqrt{<f^2>-<f>^2}}{<f>}\propto\frac{1}{\sqrt{N}}.\]
Where $N$ is the number of statistically independent units involved in the system (for example, $f$ could be the total energy of the gas and $N$ could be the total number of molecules present in the gas). In Figure (\ref{st}) we check that how $\frac{\sqrt{<N_{RR}^2>-<N_{RR}>^2}}{<N_{RR}>}$ varies with $<N_{RR}>$.
From figure (\ref{st}(d)) at $T=1491~K$ it is clear that Statistical Independence (SI) is obeyed in the disordered phase. The validity of SI in this regime is the result of the absence of long range order (various parts of the system behave statistically independently). As the temperature is lowered  we start to see deviations from statistical independence (figures \ref{st}(c) and \ref{st}(b)). In the ordered state (say at $T=301~K<T_0$) we see that SI is completely violated (Fig.(\ref{st}(a))). This clearly is the result of long range correlations (different parts of the system are statistically correlated).  Thus from this analysis it is clear that the ordered state is highly fluctuating and show long range correlations. The number of ordered pairs (short range orientational order--right-right; left-left; up-up; down-down) increase in the ordered phase which leads to the long range order.
\begin{figure}[h!]
\includegraphics[height = 5cm, width = 7cm]{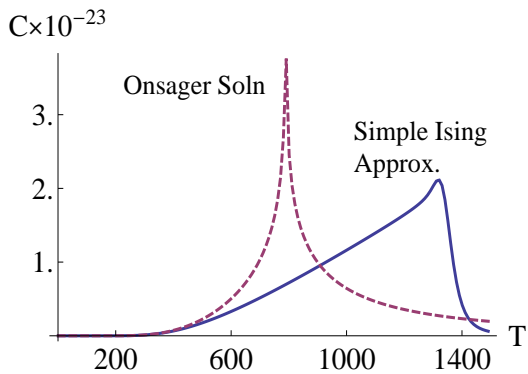}
\caption{Comparison of the reduced model with Onsager's solution.}
\label{sim}
\end{figure}
The long-range order is in the form of a spin density wave. The real space structure is something like this: two spins in each arrow (one at head and the other at tail) are roughly a lattice constant away ($~4 \AA$ in cuprates) and spins in different arrows could be many lattice spacings away (depending upon doping level). Thus there exist a fluctuating long range spin arrangement.

Further, this model reduces to simple 2-D Ising model in the following way. Let us restrict that spins can point either up or down (no left-right orientation is allowed). Then, the Hamiltonian in equation (10) will not contain terms of $N_R$ and $N_L$ and there will be no distinction (energetically) between $N_{UU}^x$ and $N_{UU}^y$, thus $\gamma=\delta=1$. The only energy left will be the ferromagnetic alignment energy $-\ep_1$, and also the sum in equation (11) will be only over $N_U$. The solution, under this reduction, reduces to the solution of the simple Ising model under Bragg-Williams approximation\cite{ker, bragg}. In Figure (\ref{sim}) we plot specific heat per arrow of this simple Ising model alongwith the result from Onsager's solution: $C_{Onsager} = \frac{2 k_b}{\pi} (\beta \ep \coth(2 \beta \ep))^2 (2 EllipticK(\kappa)-2 EllipticE(\kappa) -(1-\kappa')(\pi/2+\kappa' EllipticK(\kappa)))$ with $\kappa =\frac{2\sinh(2 \beta \ep)}{\cosh^2(2\beta\ep)}$ and $\kappa'= 2 \tanh^2(2\beta\ep)-1$. This figure is the standard result (see figure 15.6 in \cite{ker}).

Open Issues: The solution for the model presented here leaves out many issues. One can do a better job with Bethe-Peierls approximation, or still better with Onsager's like approach. Also the present solution does not apply at the critical point (as usual with mean-field solutions). A better job can be done using renormalization group study for this model.  

In conclusion: an Ising type toy model is developed in which a "spin" can point in four possible directions in a plane.  This toy model may have some analogies to the pre-formed pair picture of the pseudogap state of cuprate high temperature superconductors (however, interest here is purely mathematical). We use a slight generalization of Bragg-Williams mean-field theory to  examine the behaviour of heat capacity and an "order parameter" which capture the short range correlations between the "spins" (pre-formed pairs). Below a transition temperature these short range correlations enhance and leads to a long range order which is examined through the breakdown of statistical independence.

\end{document}